\documentclass[10pt,preprint]{iopart}
\usepackage{iopams}
\usepackage{graphicx}
\usepackage[english]{babel}
\usepackage[colorinlistoftodos]{todonotes}
\usepackage[colorlinks=true, allcolors=blue]{hyperref}
\usepackage[normalem]{ulem} 

\begin{document}
\title{Terahertz magnetic field enhancement in an asymmetric spiral metamaterial}
\author{Debanjan Polley$^1$, Nanna Zhou Hagstr\"om$^{1,2}$, Clemens von Korff Schmising$^2$, Stefan Eisebitt$^{2,3}$, Stefano Bonetti$^1$}
\address{$^1$ Department of Physics, Stockholm University, 106 91 Stockholm, Sweden}
\address{$^2$ MBI Max-Born Institute for Nonlinear Optics and Short Pulse Spectroscopy, Max-Born-Str. 2A, 
	12489 Berlin, Germany}
\address{$^3$ Institut für Optik und Atomare Physik, Technische Universität Berlin, 10623 Berlin, Germany}
\ead{stefano.bonetti@fysik.su.se}

\begin{abstract}
We use finite element simulations in both the frequency and the time-domain to study the terahertz resonance characteristics of a metamaterial (MM) comprising a spiral connected to a straight arm. The MM acts as a RLC circuit whose resonance frequency can be precisely tuned by varying the characteristic geometrical parameters of the spiral: inner and outer radius, width and number of turns. We provide a simple analytical model that uses these geometrical parameters as input to give accurate estimates of the resonance frequency. Finite element simulations show that linearly polarized terahertz radiation efficiently couples to the MM thanks to the straight arm, inducing a current in the spiral, which in turn induces a resonant magnetic field enhancement at the center of the spiral. We observe a large (approximately 20 times) and uniform (over an area of $\sim 10~\mu m^{2}$) enhancement of the magnetic field for narrowband terahertz radiation with frequency matching the resonance frequency of the MM. When a broadband, single-cycle terahertz pulse propagates towards the metamaterial, the peak magnetic field of the resulting band-passed waveform still maintains a 6-fold enhancement compared to the peak impinging field. Using existing laser-based terahertz sources, our metamaterial design allows to generate magnetic fields of the order of 2 T over a time scale of several picoseconds, enabling the investigation of non-linear ultrafast spin dynamics in table-top experiments. Furthermore, our MM can be implemented to generate intense near-field narrowband, multi-cycle electromagnetic fields to study generic ultrafast resonant terahertz dynamics in condensed matter.
\end{abstract}

\maketitle
\section{Introduction}
With the advances in the development of laser-based, table-top sources~\cite{Taniuchi2004WidelyTunable,Houard2008StrongEnhancement,Hauri2011StrongField,Wu2014StrongTHz}, strong terahertz radiation has become a novel tool to investigate low-energy excitations in condensed matter physics \cite{Hoffmann2009Kerr,kampfrath2011coherent, dienst2011bi, hu2014optically, Li2015SiWaveguide,Shalaby2017DiamondTHz,Woldegeorgis2018StrongTHz}. Within the very active research field of ultrafast magnetism, sub-ps demagnetization dynamics in metallic films triggered by single-cycle terahertz radiation have been observed~\cite{Vicario:NaturePhotonics:2013, StefanoPRL2016, shalaby2016simultaneous, Hauri2016}. Compared to the near-infrared lasers more commonly used in this type of experiments \cite{Beaurepaire1996,koopmans2000ultrafast,koopmans2005unifying, THEO-RASING-REVMODPHYS}, terahertz radiation, via its magnetic field component, can couple directly to the spin degree of freedom, rather than through thermalization of highly energetic excited states. For instance, with single-cycle terahertz magnetic fields of the order of 1 T, it is possible to induce coherent precessional or ``ballistic'' magnetization reversal in a metallic ferromagnet~\cite{stohr2007magnetism, tudosa2004ultimate, gamble2009electric}. However, it is difficult to generate such a strong terahertz magnetic field with table-top sources and, so far, only accelerator-based sources have been able to achieve a precessional magnetization reversal. 

In the last few years, the enhancement of the terahertz \emph{electric} field using metamaterials (MMs) has been realized using several methods \cite{VO2-PAPER-NATURE-KEITH-NELSON_2012,Iwaszczuk2012THzEField,Bahk2017THzEfieldSlit,Kozina2017THzXray}. However, only few works have been addressing the enhancement of the  \emph{magnetic} counterpart \cite{Mukai2014antiferromagnetic,Mukai2016Nonlinearmagnetization,Polley2018THz-driven,Qiu2018THzMagneticEnhancement}. MMs have long been studied for their exotic tunable electromagnetic response unavailable in naturally occurring materials \cite{Pendry2000Negative,Moreno2001Extraordinary,chen2008THzactiveMM,Polley2013Polarizing,PolleyTHzanti,pancaldiTHzantenna2017}. Split-ring resonators \cite{Luo2014Design, Mukai2016Nonlinearmagnetization} are one class of MMs in which an enhancement of the magnetic field due to the an eddy current circulating around the MM has been observed \cite{Padilla2006Dynamical,Kumar2012nearfield, liu2013resonance, Mukai2014antiferromagnetic, Mukai2016Nonlinearmagnetization}. While the magnetic field is indeed strongly enhanced, such an enhancement is mostly concentrated on the edges of the MM, and remains small in the central gap of the MM. Only very recently, a combination of MM and a complex three-dimensional tapered waveguide has been used to increase the terahertz magnetic field more uniformly \cite{Qiu2018THzMagneticEnhancement}. We have also recently shown that, by introducing an asymmetric arm in a circular split-ring resonator, a relatively easily fabricated structure, the terahertz magnetic field can be enhanced ten-fold in the central gap of the MM, while the corresponding electric field enhancement remained small~\cite{Polley2018THz-driven}. 

In this work, we build on our previous study by thoroughly investigating the design of an asymmetric spiral split-ring resonator that greatly enhances the magnetic component of the terahertz radiation, while by design only slightly enhancing the electric field component. We use finite element simulations in both the frequency- and the time-domain to investigate the three-dimensional distribution of electric and magnetic fields in such a structure. The MM, which can be described in terms of a resonant RLC circuit, is optimized to maximize the enhancement at 1 THz, roughly the center frequency of the 2 THz bandwidth of the radiation generated with well-established table-top laser-based sources. We also present an analytical model that can be used to calculate the resonance frequency of the MM when the inner and outer radii, the width and the number of turns of the structure are varied. Finally, we find that finite element simulations in the time-domain are necessary to reveal the complex evolution of the electromagnetic field in the MM, and to correctly estimate the peak-field enhancement with respect to an incoming broadband, single-cycle terahertz source.

\section{Metamaterial Structure}
\begin{figure*}[b]
\centering
\includegraphics[width=\textwidth]{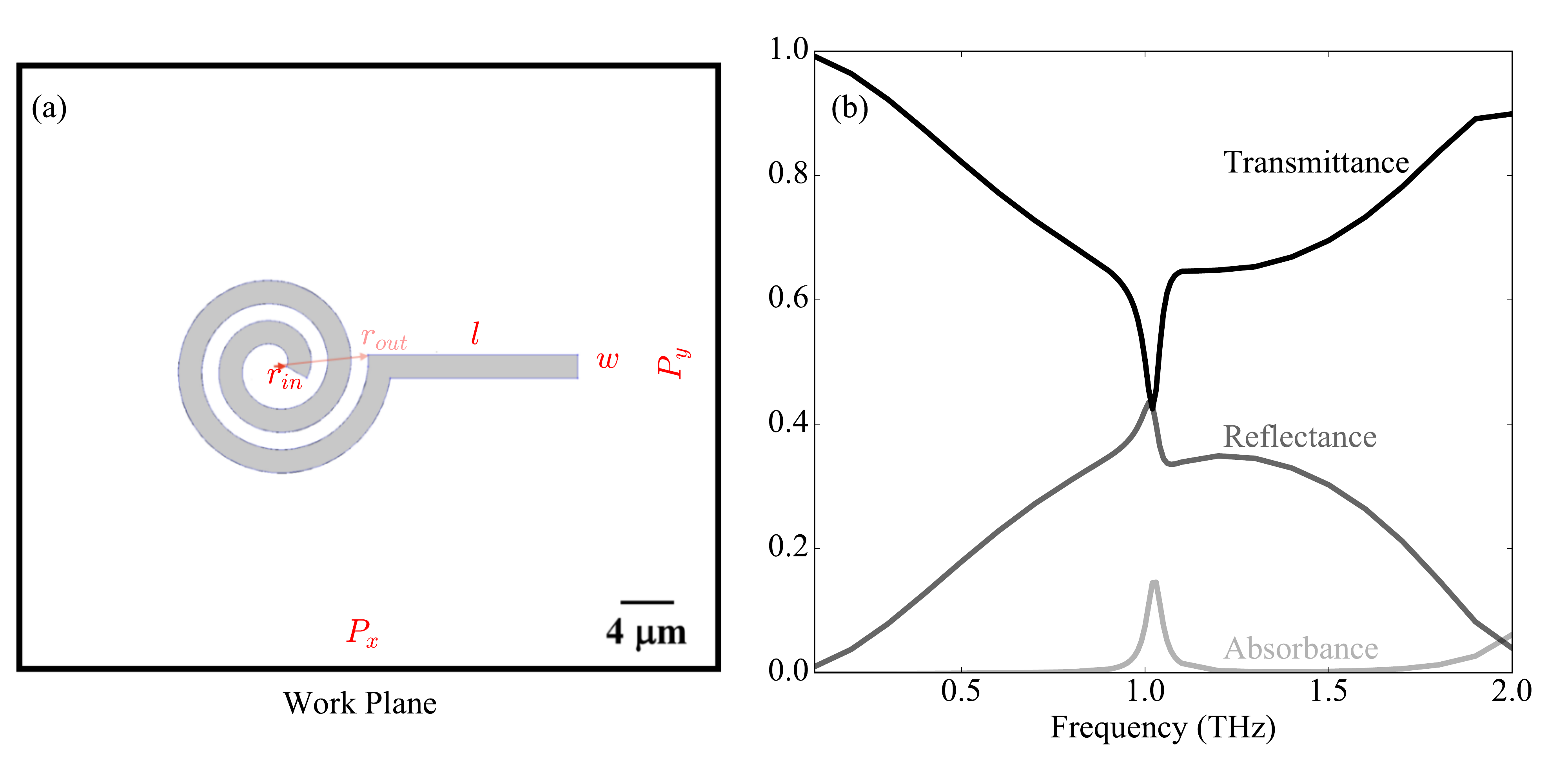}
\caption{(a) Top view of the work plane, the parameters of the MM are $P_{x}=60$ $\mu$m, $P_{y}=52$ $\mu$m, $r_{out}=8$ $\mu$m, $r_{in}=1$ $\mu$m, $n=2$, $l=18$ $\mu$m, $w=2$ $\mu$m, (b) Transmittance, reflectance and absorbance of the MM as a function of incident THz frequency.} 
\label{F1}
\end{figure*}

We use the full three-dimensional finite-element method implemented in the Wave Optics module of the COMSOL Multiphysics software \cite{COMSOL} to study the near-field enhancement of the THz electric and magnetic fields in both time and frequency domains in a resonant asymmetric spiral MM structure shown schematically in Fig.~\ref{F1}(a).

The unit cell of the MM is a $200$~nm-thick spiral asymmetric gold split-ring-resonator, on top of a $30$~$\mu$m-thick glass substrate. The structure also includes adjacent blocks of air with thickness $\lambda=300$ $\mu$m (the vacuum wavelength of electromagnetic radiation with frequency of 1 THz) on top of the gold structure, and at the bottom of the glass substrate. We used $P_{x}=60$ $\mu$m and $P_{y}=52$ $\mu$m for the lateral sizes of the unit cell. For the frequency-domain simulations, perfectly matched layers with thickness $\lambda/5$ are used at the top and the bottom boundaries of the simulation region to get rid of the unwanted reflections of the electromagnetic waves. In the time-domain simulation, the same effect is achieved implementing scattering boundary conditions. Perfect magnetic conductor (PMC) and perfect electric conductor (PEC) boundary conditions are applied to the surfaces parallel to the $yz$ and $xz$ planes, respectively, since the radiation is linearly polarized with the electric field along $x$. The gold MM has a minimum tetrahedral mesh size of approximately 13 nm. The substrate and the air blocks are divided into tetrahedral meshes with a minimum mesh size of about 0.5 $\mu$m and maximum mesh size of $\lambda/(20n)$. For the substrate, we used the experimental value $\epsilon_r=n^2=3.84$ for the relative permittivity~\cite{thzglass}. The metallic Au layer is modeled as a lossy metal with the frequency-independent conductivity of gold, $\sigma = 4.09 \times 10^7$ Sm$^{-1}$ ~\cite{AuTHzValue}.

In order to systematically control the shape of the MM, we use the parametric equation of a spiral to draw the structure as a function of four parameters: outer radius ($r_{out}$), inner radius ($r_{in}$), number of turns ($N$) and width ($w$) of the structure, all indicated in Fig.~\ref{F1}(a). Then, a rectangular arm with length $l$ and width $w$ is added to the spiral structure. The function of this straight arm is to efficiently couple linearly polarized electromagnetic radiation impinging perpendicular to the plane of the structure, and with the magnetic field perpendicular to the arm. The screening of the \emph{magnetic} field component orthogonal to the arm will in fact induce an eddy current parallel to the long side of the arm. Such current will flow inside the spiral, which will then simply acts as a coil inside which the magnetic field adds up constructively. As discussed in previous works \cite{pancaldiTHzantenna2017, Polley2018THz-driven}, we stress once more that the electric field component is screened very effectively at the gold surface. The typical terahertz electric fields generated with table-top sources (up to 100 MV/m, 1 ps in duration) can in fact be very effectively screened by a slight rearrangement of the surface charges in the metal. These charges can respond on the femtosecond time scale, and are subjected to interatomic electric fields more than two order of magnitude larger than the impinging terahertz field. Hence, almost all electromagnetic energy inside the conductor is in the magnetic field component of the radiation \cite{Jackson}.

\section{Resonance Frequency}

We performed full three-dimensional finite element simulations in the frequency domain in the range $0.1-2.0$ THz. Near the resonance frequency, between 0.9 THz and 1.1 THz the frequency step is set to 0.01 THz, while in the rest of the frequency range we used a frequency resolution of 0.1 THz. The present MM structure is optimized for maximum terahertz field enhancement around 1.0 THz, which is in proximity of the center of the bandwidth of commonly used broadband table-top THz sources~\cite{  HoffmanLiNBO3_2007, OH12008, Seo2009Terahertz, hoffmann2011coherent, jepsen2011terahertz, PolleyTHzanti}.
For the geometrical parameters given in the caption of Fig.~\ref{F1}, the transmittance ($T$), reflectance ($R$), and absorbance ($A$) exhibit a sharp feature at 1.02 THz, as shown in Fig. ~\ref{F1}(b). The absorption exhibits a sharp peak with the maximum value of $\approx16\%$, while the broadband, bell-shaped reflectance spectra is slightly perturbed, reaching a maximum value of $\approx42\%$. The transmittance varies also slowly, and dips to $\approx42\%$ at the resonance frequency.

\begin{figure}[b]
\centering
\includegraphics[width=\textwidth]{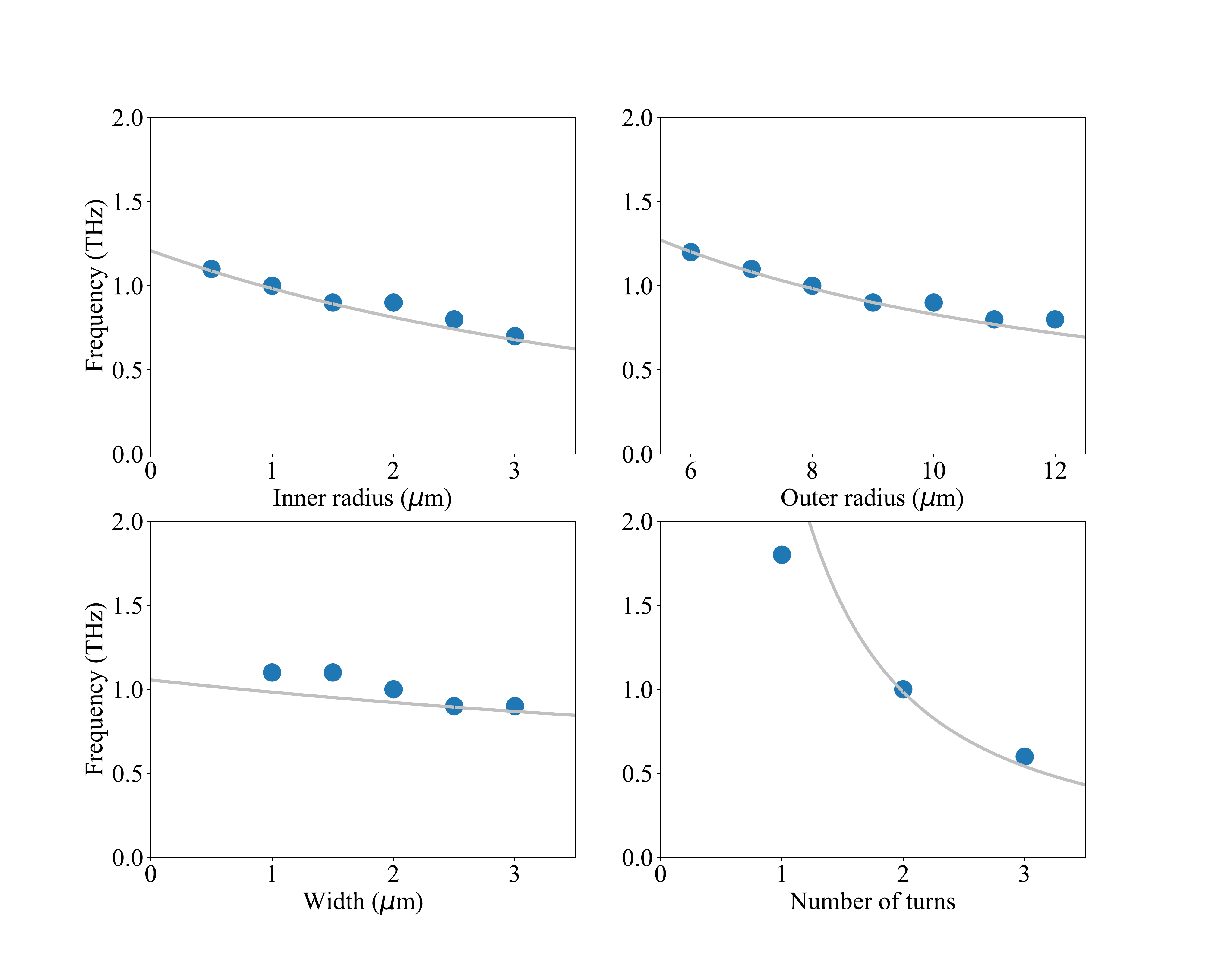}
\caption{The dependence of the resonance frequency on the structure parameters of the MM namely (a) inner radius ($r_{in}$), (b) outer radius $r_{out}$, (c) width ($w$) and (d) number of turns ($N$) of the metamaterial. The continuous line is our theoretically prediction considering the MM as a resonant RLC circuit.}
\label{F2}
\end{figure}

By varying the MM geometrical parameters, the resonance frequency can be tuned~\cite{Koschny2003Resonant,Baena2004Artificialmagnetic, Johnson2006Areview}. We have studied such frequency tuning, specifically its dependence on the four variables that parametrize the spiral. We varied each one of these variables, time while keeping the other three fixed at the values reported in the caption of Fig.~\ref{F1}. The solid symbols plotted in the four panels of Fig.~\ref{F2} show the simulated resonance frequency of the MM as a function of the variable reported in the $x-$axis. The general trend is the decrease of resonance frequency with increasing structure size. For these plots, the reported frequency is correct within 0.1 THz, the step size of the finite element numerical simulations.

While these numerical simulations are very accurate in predicting the correct resonance frequency and enhancement, they are also computationally very demanding. Each simulated point took $\sim$ 10 hours on a 40-core workstation with 128 GB of RAM, making the design from scratch of an arbitrary structure unpractical. For this reason, we explored the possibility of using an analytical model to describe the dependence of the resonance frequency on the four geometrical variables. A resonant MM can in general be described as a RLC circuit where the $R$, $L$ and $C$ are the inductance, capacitance and resistance, respectively, of an equivalent circuit model of the structure. The resonance frequency of such a circuit would then be $f = 1/2 \pi \sqrt{LC}$, where the inductance, the capacitance of a spiral are approximated with the expressions valid for concentric coils for L \cite{spiral_coil}, and for a solenoid for C \cite{capacitance_solenoid}:
\begin{eqnarray}
L = \frac{\mu_{r}\mu_{0}} {2\pi}\left[\frac{(ND/2)^{2}}{100(2D+5.3(r_{out}-r_{in}))}\right],\label{eq:L}\\
C = \frac{4\epsilon_{r}\epsilon_{0}}{\pi}l\left[1+0.71\left(\frac{D}{l}\right)+2.40\left(\frac{D}{l}\right)^\frac{3}{2}\right]\label{eq:C},
\end{eqnarray}
where $\mu_{0} $ and $\epsilon_{0}$ are the vacuum permeability and permittivity, respectively,  and, $\mu_{r}$ and $\epsilon_{r}$ the relative permeability and permittivity, respectively, of the effective medium.  $D$ and $l$ are the average diameter and total length of the spiral, respectively. The length of an Archimedean spiral is
\begin{equation}
l = \int_{0}^{2\pi N} \sqrt{r^{2}+\left(\frac{dr}{d\theta}\right)^{2}} d\theta, 
\end{equation}
where $r$ is the radius and $N$ the number of turns of the spiral.

By scaling the resonance frequency with a single parameter $c$, we rewrite the equation for the resonance frequency of the spiral as
\begin{equation}
f_{s} = \frac{c}{2 \pi \sqrt{LC}}.
\label{eq1}
\end{equation}
The predicted resonance frequencies for the analytical model are plotted as the gray solid line in Fig.~\ref{F2}. Assuming the single value $c = 1.4$, the analytical equations reproduce well the resonance frequency of the metamaterial as a function of each of the four geometrical variables. The correction a factor $c$ is to be expected, as our spiral is only an approximation of the concentric coil or of the solenoid modeled by Eqs. (\ref{eq:L}) and (\ref{eq:C}). This discrepancy is particular evident for the case when the number of turns $N=1$, i.e. when the approximation of the coil with a solenoid is least valid.

The resistance of the MM can be computed in a straightforward way as $R = \rho{l}/{wt}$, where $\rho$ is the resistivity of the metal used for the metamaterial, and $t$ its thickness. The bandwidth of the resonance is then $\Delta f=2\pi R/L$, which returns the correct order of magnitude $\Delta f\approx0.1$ THz for the parameters given in the caption of Fig.~\ref{F1}.

\section{Field Enhancement}

\begin{figure}
\centering
\includegraphics[width=\textwidth]{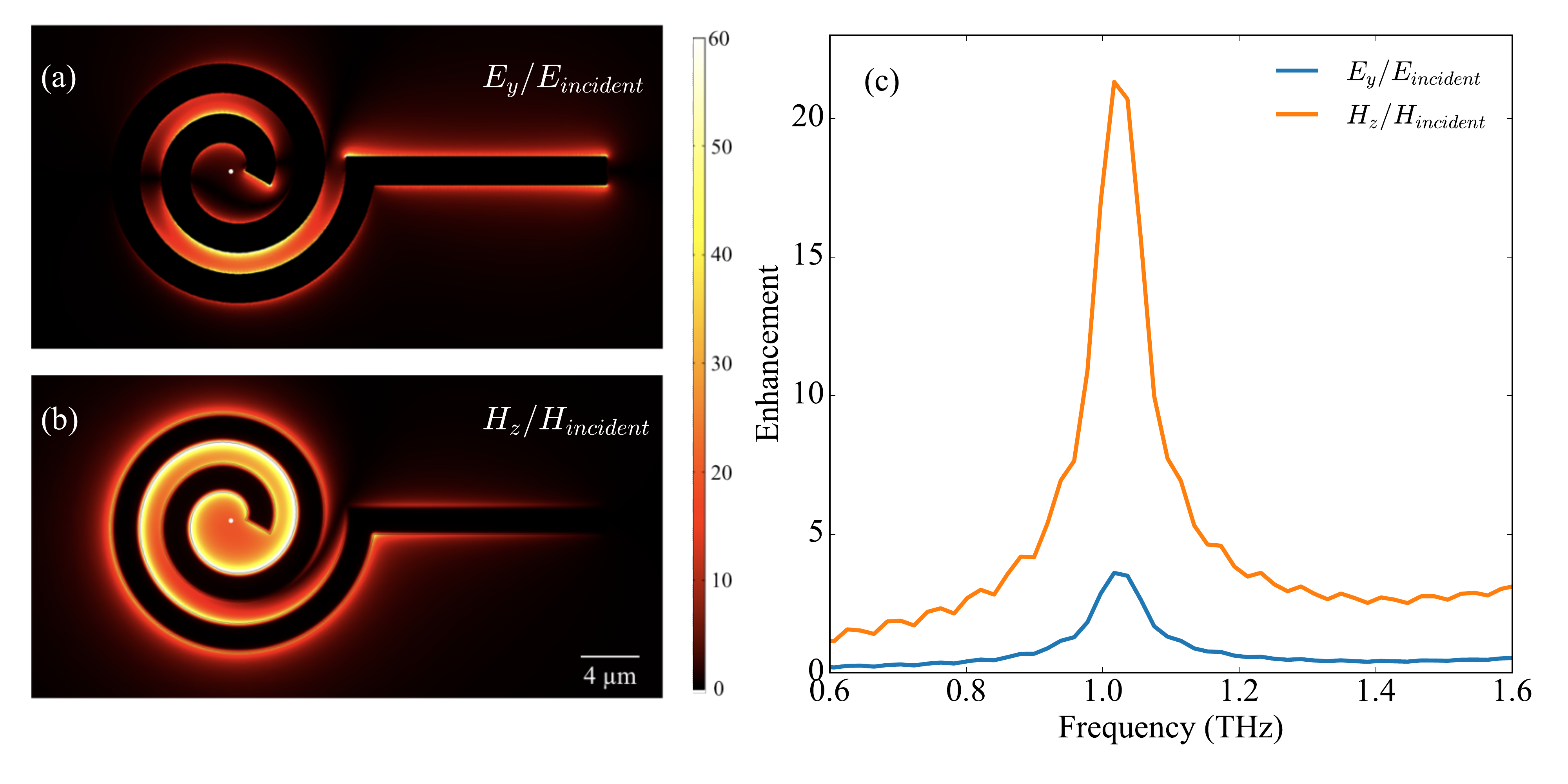}
\caption{The (a) electric and (b) magnetic field enhancement (with respect to the incident field) at the MM plane observed from the frequency domain analysis. The enhancement observed by Fourier transforming the time domain data from Fig. \ref{F4} is shown in (c). The white dot in (a) and (b) is the origin of the spiral, where we have calculated the enhancement in the time domain simulation.}
\label{F3}
\end{figure}

In order to compute the enhancement of the electric and magnetic field components of the terahertz radiation, we performed finite element simulations both in the frequency and the time domain: the frequency domain results allow to draw a two-dimensional ``enhancement map'' in the region surrounding the MM; the time-domain computations return the detailed temporal evolution of the electric and magnetic fields at a given point in space. The time-domain simulations are particularly relevant for experiments where broadband, single-cycle terahertz pulses are used to excite dynamics in materials. In the following, we assume a MM structure with the dimensions reported in the caption of Fig.~\ref{F1}

For the frequency domain analysis, we have used an approach similar to our earlier work to calculate the enhancement factor~\cite{Polley2018THz-driven}. Briefly, the terahertz electric $E_{incident}$ and magnetic $H_{incident}$ fields (assumed to be uniform over the sample plane) incident on the metamaterial are taken as reference. Once the terahertz electric $E_{y}$ and magnetic $H_{z}$ field values at the MM surface are determined, they are normalized by the incident fields to determine the enhancement factor. We chose the $y$ and $z$ component of the electric and magnetic fields, respectively, as they are the ones most strongly enhanced by the MM structure discussed here.

The enhancement maps for the electric and the magnetic fields at the MM surface are the plotted in Fig.~\ref{F3}(a) and, respectively, Fig.~\ref{F3}(b). The magnetic field enhancement is moderately uniform over the central gap, and it is clearly larger than the electric field enhancement. Furthermore, the electric field is mostly enhanced in the outer annular region and in the outer edge of the spiral MM, and it is only slightly enhanced in the central gap of the spiral. This is clearly shown in Fig.~\ref{F3}(c), where we plot the enhancement factor at the center of the spiral in the region indicated by the white dot in Fig.~\ref{F3}(a)-(b). The curves in Fig.~\ref{F3}(c) are the Fourier transform of the simulated time-domain traces that will be discussed below. It can be clearly seen that the electric field enhancement is a factor of about 3, while the magnetic field enhancement exceeds a factor of 20. 

\begin{figure}
\centering
\includegraphics[width=\textwidth]{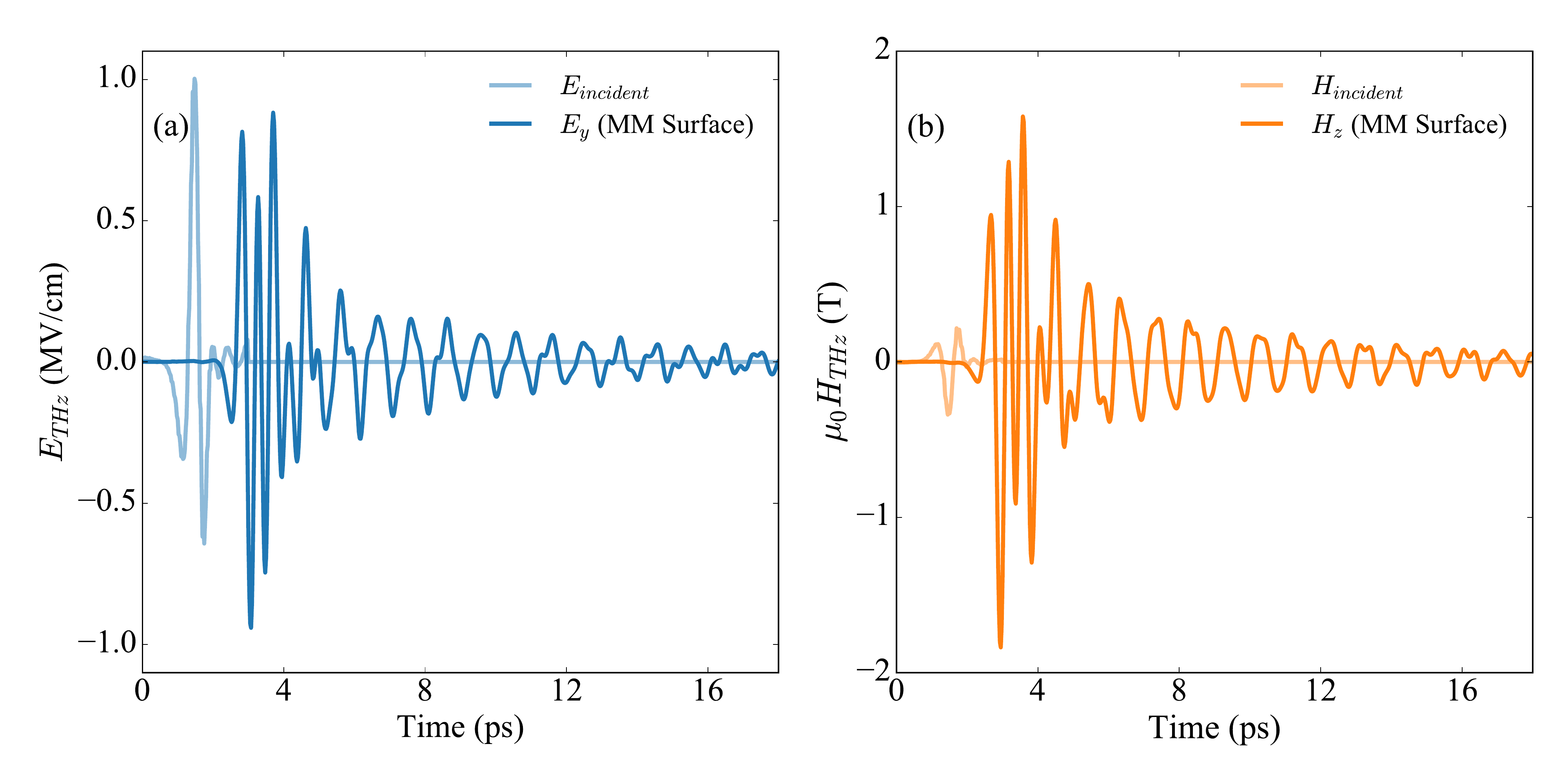}
\caption{The (a) electric ($E_{y}$) and (b) magnetic field ($H_{z}$) profile at the origin of the spiral MM as obtained from the time-domain simulation. The incident electric $E_{incident}$ and magnetic $H_{incident}$ fields are shown in semitransparent color.}
\label{F4}
\end{figure}

We now turn to the time-domain simulations. In this case, we used as the reference field the electric field of a typical broadband, single-cycle terahertz pulse measured experimentally using electro-optical sampling in a GaP crystal. The terahertz electric field is plotted as the semi-opaque line in Fig.~\ref{F4}(a), with a peak of approximately 1 MV/cm. The corresponding peak magnetic field has a value of 0.33 T. These are field values that are nowadays attainable with state-of-the-art laser-based terahertz sources. In Fig.~\ref{F4}(a) and (b) we plot the calculated enhancement of the $y$ component of the electric field and, respectively, of the $z$ component of the magnetic field at the center of the spiral. The most striking feature for both field components is that the single-cycle, broadband incident field has now the shape of a damped narrowband oscillation at a frequency of approximately 1 THz. This is not surprising considering the narrow resonance of the structure plotted in Fig.~\ref{F1}(b). It is however worth highlighting this fact, as the broadband characteristics of the pulse can sometimes be preserved in different types of MM structures, when the resonance frequency is far off the bandwidth of the impinging radiation \cite{pancaldiTHzantenna2017, pancaldi2018anti}.

Considering the field enhancement, the MM structure does not substantially modify the peak value of the electric field component, as compared to the incident field and as shown in Fig.~\ref{F4}(a). On the other hand, the magnetic field component of the radiation is enhanced by a factor of roughly 6, reaching peak values close to 2 T. The apparent discrepancy in the simulated enhancement factor between the frequency- and the time-domain simulations is readily explained. The peak-field enhancement in the time-domain is the overall, broadband enhancement. Conversely, the frequency-domain analysis returns the enhancement factor at a single frequency. In order to confirm the consistency between the two simulations in the reciprocal domains, we note that the expected enhancement at the resonance frequency is the same for both frequency domain simulations plotted in Fig.~\ref{F3}(a)-(b), and the one computed by Fourier transform of the time-domain trace, shown in Fig.~\ref{F3}(c).

Finally, the time-domain simulations  demonstrate that our MM structure can be used not only to enhance the magnetic field component of incident terahertz radiation, but more generally to generate intense narrowband, multi-cycle terahertz pulses starting from broadband ones. While broadband terahertz radiation is nowadays commonly generated with table-top lasers, narrowband radiation is still relatively challenging to achieve in these systems \cite{lee2000generation, shi2003continuously, chen2011generation, carbajo2015efficient}, and accelerator-based sources are needed \cite{stojanovic2013accelerator, green2016high}. With our MM design, intense electric (1 MV/cm) and magnetic (2 T) narrowband terahertz fields become easily accessible in small-scale laboratories.

\section{Conclusion}

We have used finite element method simulations in both the frequency- and time-domain to calculate numerically the near-field enhancement in a planar, asymmetric spiral metamaterial structure. We designed the MM to provide resonant enhancement at around 1 THz, roughly the center frequency of the broadband terahertz radiation generated with state-of-the-art laser-based sources. We have shown that the dependence of the resonance frequency on the geometrical parameters of the structure can be accurately predicted by a simple analytical model that describes the MM structure as an approximate resonant RLC circuit.

Numerical simulations have shown large enhancement of the terahertz magnetic field (20-fold at the resonant frequency, 6-fold over the broadband), and negligible electric field enhancement (3-fold at the resonant frequency, no enhancement over the broadband). This is desirable, as strong electric fields can cause dielectric breakdown and be potentially damaging for solid state experiments. The magnetic field enhancement is quite uniform over a $\sim 10$ $\mu$m$^{2}$ region at the center of the spiral. We envision that such MM will open up new possibilities for experiments of ultrafast nonlinear spin dynamics in different magnetic systems.

Additionally, our MM can be used as a table-top near-field source of narrowband, multicycle terahertz radiation driven by broadband terahertz pules naturally generated by laser-based sources. The peak fields achievable (electric field: 1 MV/cm, magnetic field: 2T) are large enough to induce nonlinear effects in solids. We anticipate that our simple MM design will enable the investigation of resonant and nonlinear terahertz dynamics of electrons, phonons and magnons in condensed matter experiments.

\section*{Acknowledgments}
We are grateful to Matteo Pancaldi for the support with COMSOL simulations, and to Paolo Vavassori for fruitful discussion on the interaction between matter and electromagnetic radiation. DP and SB acknowledges support from the European Research Council, Starting Grant 715452 'MAGNETIC-SPEED-LIMIT'.

\section*{References}
\bibliographystyle{unsrt}
\bibliography{Pap1_ref}

\end{document}